\documentclass[a4paper,11pt]{article}
\pdfoutput=1 

\usepackage{jinstpub} 
\usepackage{url}

\newcommand{\JSNS}{JSNS$^2$\,}
\newcommand{\dMsq}{$\Delta m^2$\,}

\newcommand{\nuebar}{$\bar{\nu}_{e}$}

\newcommand{\Sterile}{$\bar{\nu}_{\mu} \to \bar{\nu}_{e}$~}

\newcommand{\IBD}{$\bar{\nu}_{e} + p \to e^{+} + n$~}

\newcommand{\mcub}{$\mathrm{m}^3$\,}
\newcommand{\msq}{$\mathrm{m}^2$\,}
\newcommand{\micro}{$\mu \mathrm{s}$\,}

\newcommand{\eVsq}{$\mathrm{eV}^2$}
\newcommand{\Hz}{$\mathrm{Hz}$\,}

\newcommand{\mssq}{$\mathrm{m/sec^2}$\,}
\newcommand{\degC}{$^{\circ}\mathrm{C}$\,}
\newcommand{\diff}{\mathrm{d}}
\title{Stainless steel tank production and tests for the \JSNS neutrino detector}

\author[a,b,1]{Y.~Hino,\note{\, Corresponding author.}}
\author[c]{H.~Furuta,}
\author[b]{S.~Hasegawa,}
\author[c]{T.~Maruyama,}
\author[c,2]{K.~Nishikawa,\note{\, Deceased}}
\author[c]{J.S.~Park,}
\author[a]{F.~Suekane,}
\author[d]{and Y.~Sugaya}

\affiliation[a]{Research Center for Neutrino Science, Tohoku University,\\6-3 Aramaki Aza-aoba, Aoba-ku, Sendai, Miyagi, Japan}
\affiliation[b]{Advanced Science Research Center, JAEA,\\2-4 Shirakata, Tokai, Ibaraki, Japan}
\affiliation[c]{High Energy Accelerator Research Organization (KEK),\\1-1 Oho, Tsukuba, Ibaraki, Japan}
\affiliation[d]{Research Center for Nuclear Physics, Osaka University,\\10-1 Mihogaoka, Ibaragi, Osaka, Japan}

\emailAdd{hino@awa.tohoku.ac.jp}

\abstract{
This paper describes the design and the construction of the stainless steel tank of the \JSNS detector. 
The leakage was examined using water and gas after the construction. The new sealing technique with liquid gasket was developed, and its sealing capability was evaluated quantitatively. The result shows over 5 times better value than the tolerance level of leakage.
The acceleration measurement during the transportation of the tank shows adequate robustness.
These tests prove that the stainless steel tank is feasible to use the real experiment.
}


\proceeding{Preprint typeset in JINST style}

\begin{document}
\maketitle
\flushbottom

\section{Introduction}
\begin{figure}[htbp]
    \centering 
    \includegraphics[width=.6\textwidth]{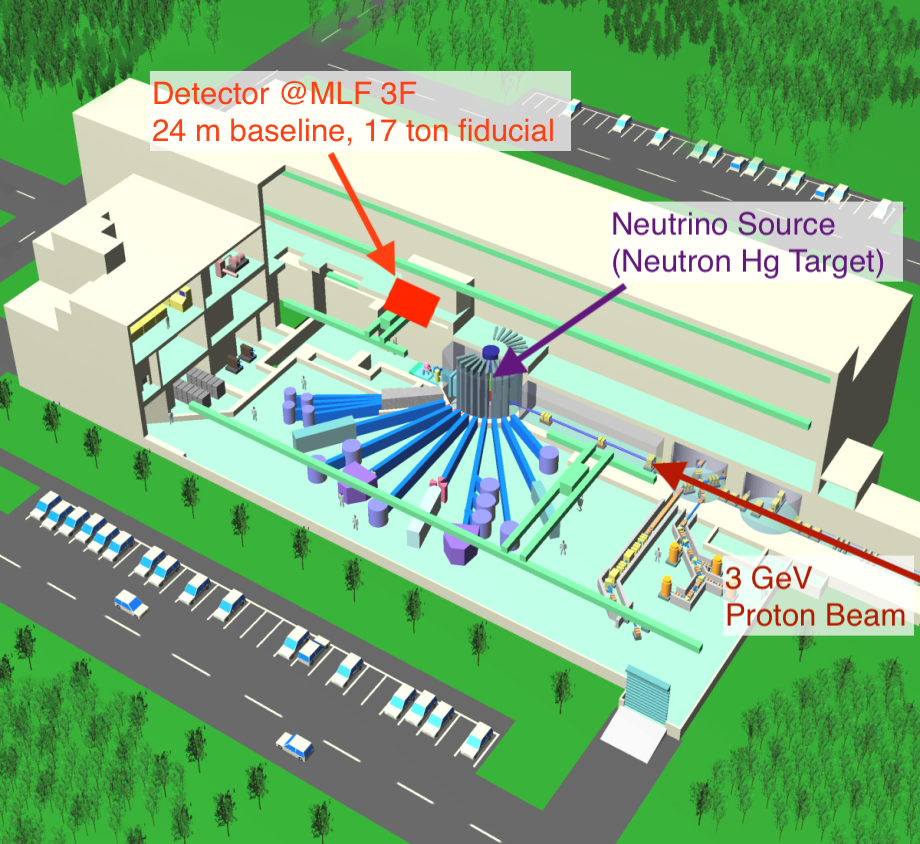}
    \caption{A bird's-eye view of J-PARC MLF and the \JSNS detector location. The red rectangle in the figure indicates the place where the (anti-)neutrino detector is settled. \label{fig:MLF_nu}}
\end{figure}

\JSNS (J-PARC Sterile Neutrino Search at J-PARC Spallation Neutron Source) \cite{Harada:2013} is an experiment to search for neutrino oscillations with \dMsq $\sim 1$ \eVsq, which is reported by LSND experiment with 3.8 $\sigma$ sensitivity in 1998 \cite{Aguilar:2001}, via the observation of \Sterile appearance oscillation. 
The experimental setup of \JSNS consists of (anti-)neutrino detector placed 24 m away from the mercury target in Materials and Life Science Experimental Facility (MLF) of J-PARC as shown in figure \ref{fig:MLF_nu}. The detector contains 17 tons of Gadolinium (Gd) loaded liquid scintillator in a neutrino target (NT) volume to detect \nuebar \, via the inverse beta-decay (IBD) reaction \IBD in a delayed coincidence method. The positron yields scintillation instantaneously which is detected as a prompt signal. When the neutron after thermalization is captured by Gd, several gamma-rays which have around 8 MeV in total are emitted by Gd. The gamma-rays generate scintillation observed as a delayed signal around 30 \micro behind the prompt signal. The delayed coincidence of the prompt and the delayed signals identifies \nuebar \, signal.

\begin{figure}[htbp]
    \centering 
    \includegraphics[width=.7\textwidth]{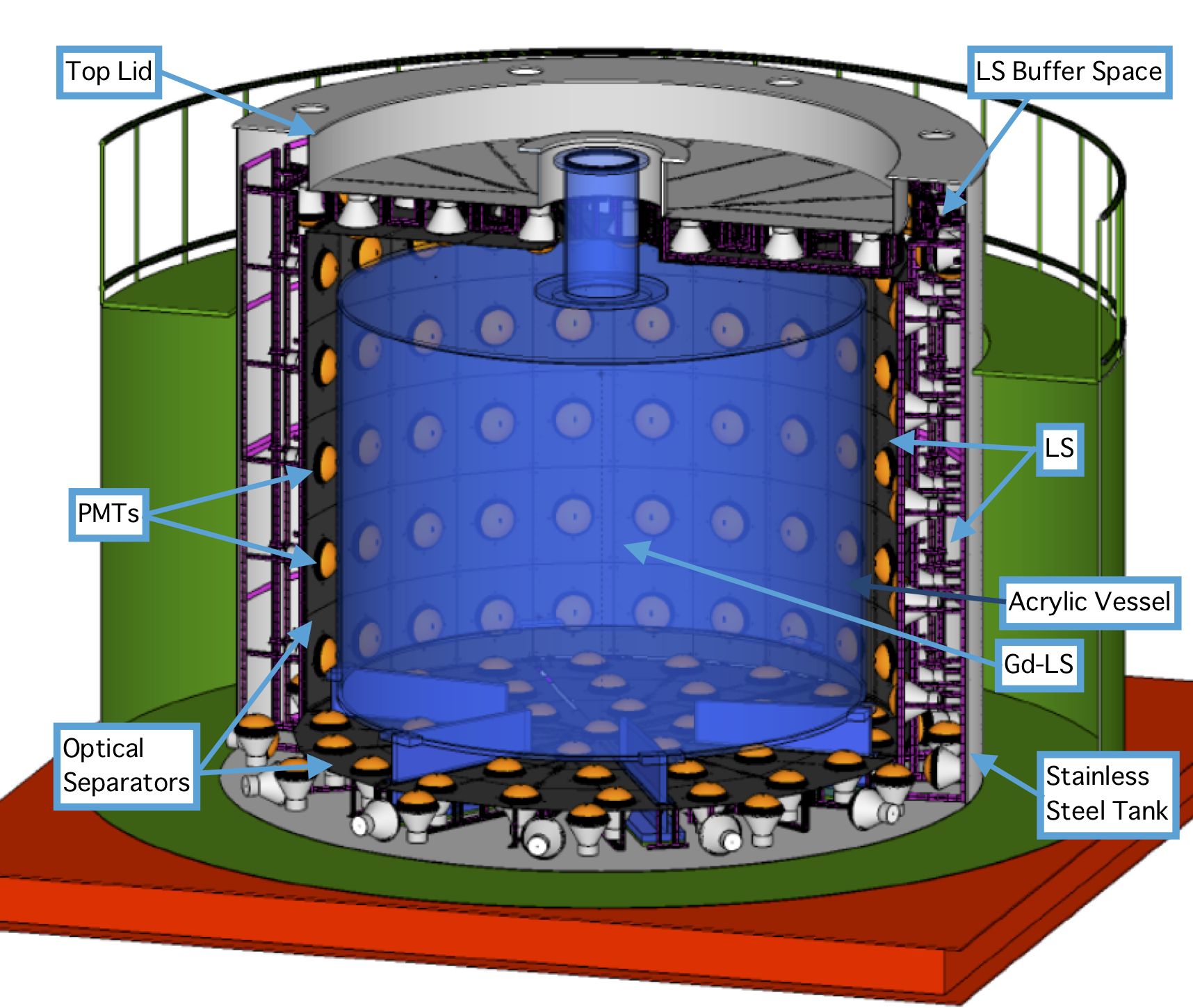}
    \caption{A cutaway view of a detailed 3D model of the \JSNS detector. The blue cylindrical container placed at the center of the detector is the acrylic vessel for Gd-LS. PMTs and black boards surrounds the vessel. Parts of the support structures for PMTs, colored pink, are welded to the inner surface of the stainless steel tank. \label{fig:Detector}}
\end{figure}

The \JSNS detector is composed of three layers of two types of liquid scintillator in a stainless steel tank whose volume is around 60 \mcub (shown in figure \ref{fig:Detector}). The most-inner of the detector has an ultraviolet light transparent acrylic vessel to contain Gd loaded liquid scintillator (Gd-LS) for detecting \nuebar. The space between the stainless steel tank and the acrylic vessel is filled with Gd unloaded liquid scintillator (LS), which is optically separated into two layers by black boards (optical separators). The inner layer is a gamma catcher (GC) to absorb energy of gamma-rays from Gd. Inner photomultiplier tubes (PMTs) are attached on the black boards to observe lights from both NT and GC. The outer LS layer has a role of cosmic ray anti-counter. 

In addition to the delayed coincidence technique for the IBD detection, the \JSNS detector has a pulse shape discrimination (PSD) capability as a particle identification of neutral particles, especially cosmic ray induced fast neutron and neutrino signal, as described in \cite{Ajimura:2017}. In general, it is crucial to remove dissolved oxygen properly and maintain an environment preventing oxygen contamination for keeping optical properties, such as a light yield and a PSD capability, of liquid scintillator in terms of oxygen quenching effect.

This paper describes the stainless steel tank and its related tests including liquid leakage and gas-tightness of the tank. We developed a new sealing scheme for a large flange with a poor flatness, a quantitative measurement technique and evaluation method of gas-tightness using decrease of a relative pressure. They are described in detail for the future work or similar type of detectors, e.g., reactor neutrino monitors.

\section{Tank Design and Structure}
The detailed structure of the \JSNS detector is explained elsewhere \cite{Ajimura:2017}. Therefore, this section concentrates on the design and the structure of the stainless steel tank. A detailed drawings were developed by Morimatsu Industrial Co. Ltd., based on a conceptual design from \JSNS collaborators \cite{web:Morimatsu}.

\subsection{Main Tank, Top Lid and Anti Oil-Leak Tank Design}
\begin{figure}[htbp]
    \centering 
    \includegraphics[width=.8\textwidth]{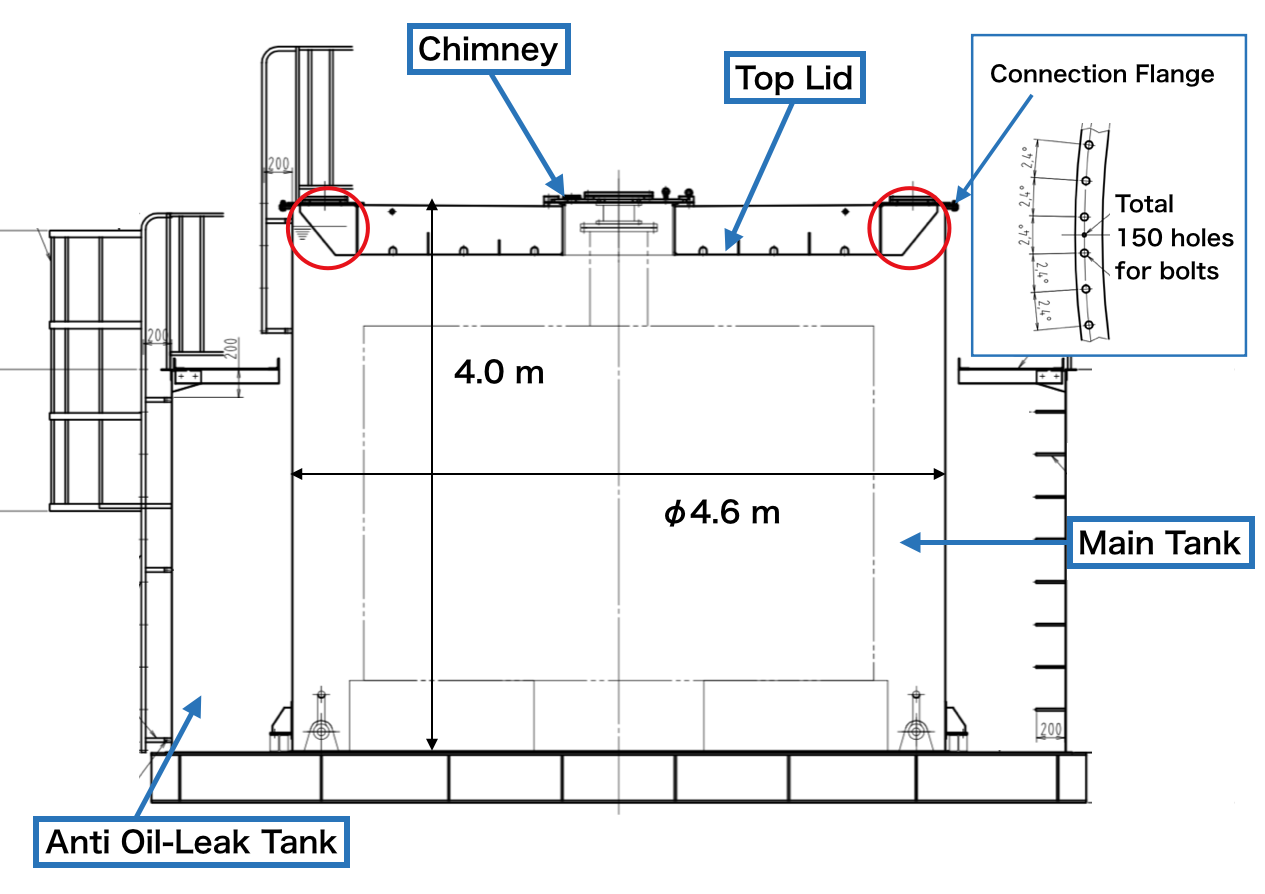}
    \caption{A side view of the tank and the anti oil-leak tank. The red circles correspond to the LS buffer space, which forms a large ring shape, to prevent a change of liquid level due to thermal expansion of LS. \label{fig:Drawings}}
\end{figure}

\begin{figure}[htbp]
    \centering 
    \includegraphics[width=.7\textwidth]{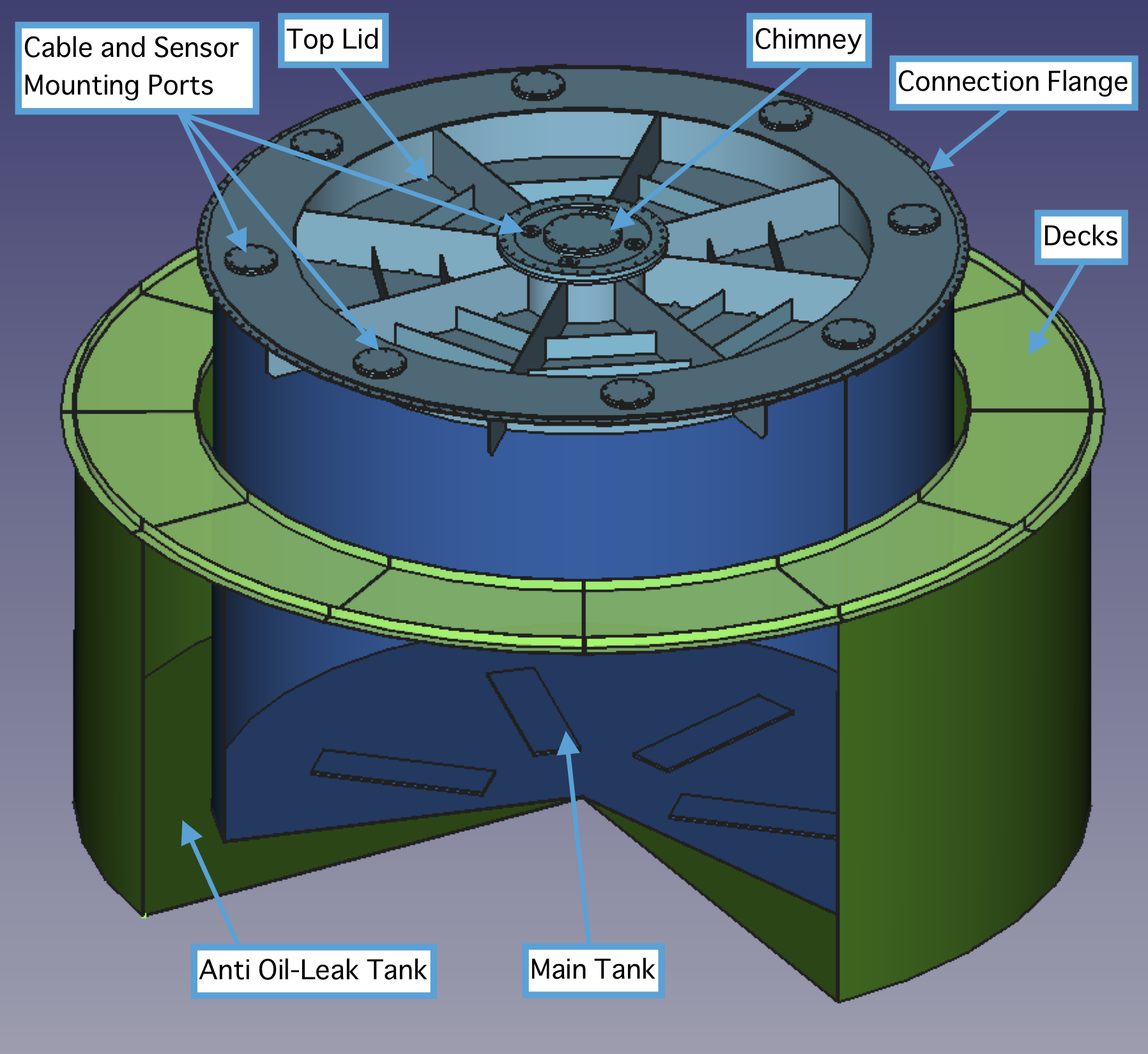}
    \caption{A detailed 3D model of the drawing without balustrades on the decks. The anti oil-leak tank is shown as a cutaway of the green cylinder with decks for operators to step on on the top. A cutaway view of the dark blue object corresponds to the main tank of the stainless steel tank. The top lid colored light blue has a large flange structure around the circumference for connection onto the main tank. \label{fig:3DModel}}
\end{figure}

The stainless steel tank of the \JSNS detector consists of two parts; a main tank and a top lid. There is an anti oil-leak tank surrounding the tank, which prevents liquid from spreading out in case of leak from the main tank. The side view drawings and the detailed 3D model of the entire tank are shown in figure \ref{fig:Drawings} and figure \ref{fig:3DModel} respectively.

The main tank is a cylindrical shape with dimension of 4.6 m diameter and 4.4 m height. The thickness of the stainless steel is 5 mm. On the top of the main tank, there is a large flange structure along the circumference pointed in figure \ref{fig:3DModel} as the connection flange to prevent gas or liquid leakage from the main tank, where the top lid resides through a sealing material.

As indicated in figure \ref{fig:Detector} and \ref{fig:Drawings}, the top lid forms the LS buffer space to absorb thermal expansion of LS along the circumference of the tank. The base area of this LS buffer ring is approximately 4 \msq, and corresponds to a capability of absorbing the LS level change within $\pm 10$ cm in $\pm 10$ \degC temperature change.

The top surface of the buffer ring has eight flange ports, which are used for a feed-through of PMT cables, mounting LS filling pipes, and ports of a nitrogen purging system. 
There are eight reinforce beams along the radial direction, and sixteen reinforce plates in the polar direction on the top-lid.
Thanks to this reinforce structure, the stainless steel tank can tolerate up to 0.2 atm relative pressure with respect to the outside pressure.

\subsection{Sealing around the connection flange}
To compensate a poor flatness of the connection flange, we decided to use Herme-seal No.800, which is a liquid type gasket with oil-proof provided by Nihon Hermetics Co. Ltd., as the sealing material instead of a o-ring or a rubber gasket \cite{web:Hermetics}. A merit of liquid type gaskets is a capability of filling gaps caused by the poor flatness.

Herme-seal No.800 loses elasticity as it gets dry. Thus, it is necessary to avoid desiccation before the lid closure is done. This can be done with mixing 10 w\% of a water dominant special diluent produced by the company into the liquid type gasket. As a result of it, the desication time was extended into more than 30 min, which is enough for our purpose.

\begin{figure}[htbp]
    \centering 
    \includegraphics[width=.5\textwidth]{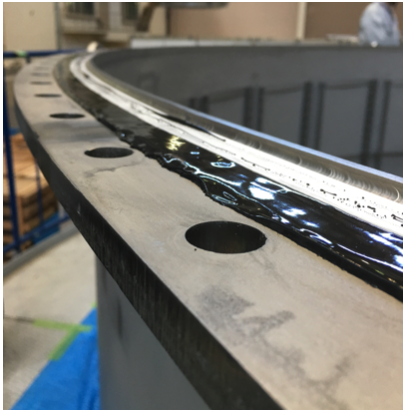}
    \caption{A photo showing the situation when Herme-seal was heaped on the connection flange. \label{fig:SealPhoto}}
\end{figure}

For the actual sealing work, we put 3 mm thick of the diluted Herme-seal on the flange as shown in figure \ref{fig:SealPhoto}. The lid was placed to close as soon after the painting work as possible. A sealing capability for gas (gas tightness) was examined after this work, which is described in section~\ref{sec:gasleaktest}.

\section{Construction}
The construction of the stainless steel tank began on December 2017, and finished in the end of February 2018. Morimatsu was in charge of construction of the tank. All construction processes were done in J-PARC in the open air. The production of the components was done at the factory of Morimatsu in advance.

\subsection{Liquid Leak Check}
\begin{figure}[htbp]
    \centering 
    \includegraphics[width=.8\textwidth]{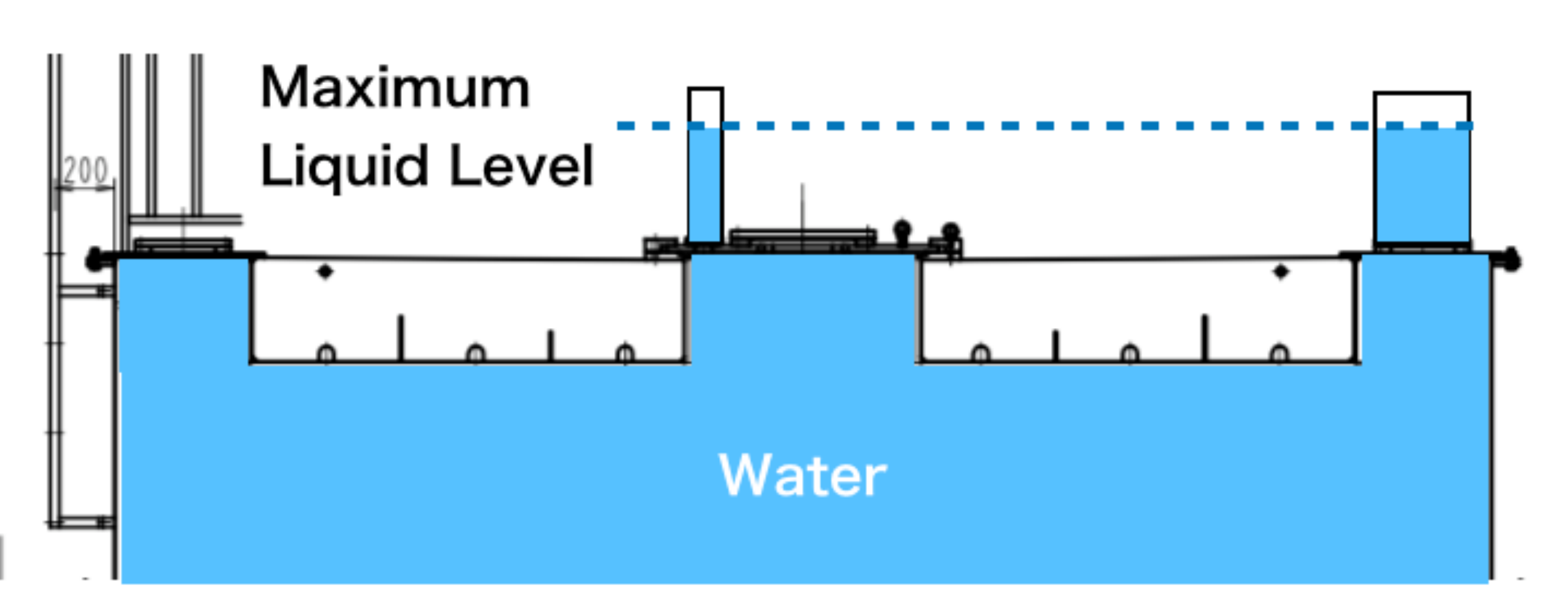}
    \caption{A schematic of a liquid level during the liquid leak test. \label{fig:FigLiqLeak}}
\end{figure}

After construction of the main tank and the top lid, we filled the tank with water to check a liquid leakage from the welded joints and flanges. The liquid level was set above the flanges of the top lid as shown in figure \ref{fig:FigLiqLeak}. To see the leakage, we left the tank in this situation over one night, and then searched the leakage points and checked the liquid level. As a result, no serious liquid leakage was found.

\subsection{Transportation in J-PARC}
The tank was transported from the construction place (M1) to HENDEL building for the storage and the further construction mainly dedicated for works inside of the tank, such as a PMT mantling. As described in \cite{Ajimura:2017}, the empty \JSNS detector is planned to be transported between HENDEL and MLF every summer. Therefore, this tank transportation will be a simulation for the planned detector transportation.

\begin{figure}[htbp]
    \centering 
    \includegraphics[width=.7\textwidth]{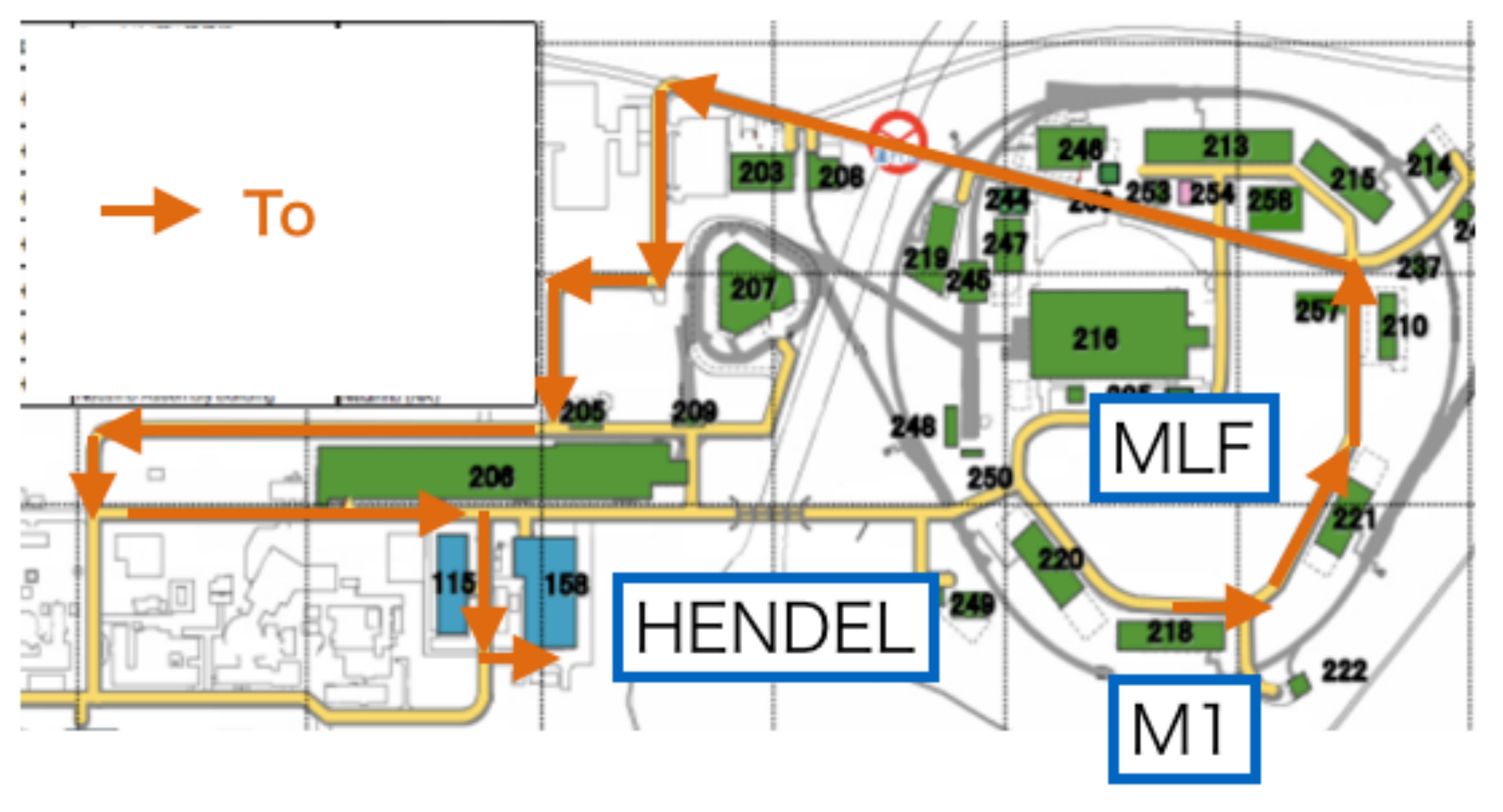}
    \caption{The route to transport the stainless steel tank from M1 (the construction place) to HENDEL building (the storage building). \label{fig:TransMap}}
\end{figure}

\begin{figure}[htbp]
    \centering 
    \includegraphics[width=.9\textwidth]{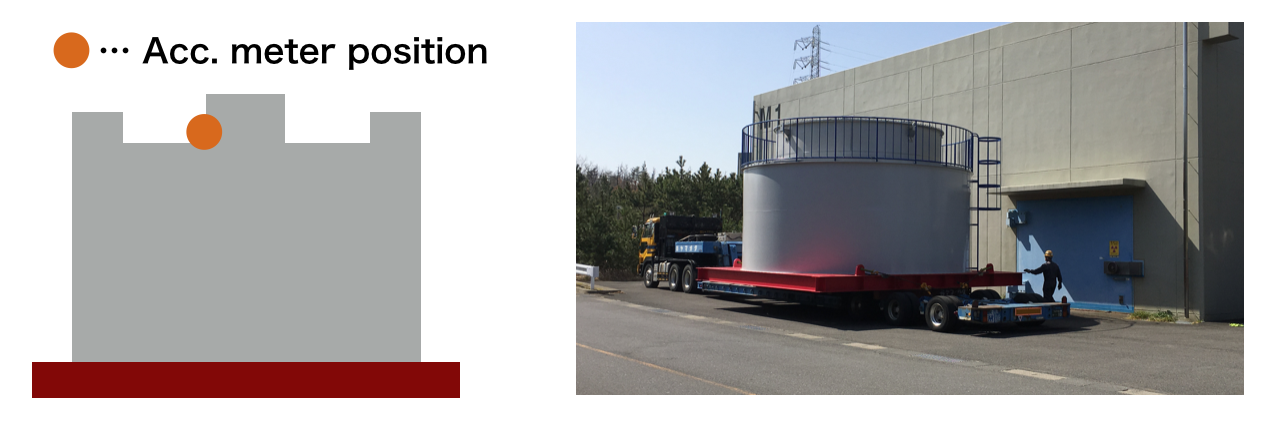}
    \caption{Left: the position of the acceleration sensor. Right: a photo of the tank and the low bed trailer truck. \label{fig:TransPhoto}}
\end{figure}

The orange arrows in figure \ref{fig:TransMap} show the course for the tank transportation towards HENDEL building, which is the same way as the actual detector transportation after the junction to MLF. The tank was transported by a low bed trailer truck (shown in the right of figure \ref{fig:TransPhoto}). The truck went along the course slowly (less than 20 km/h) for safety. We attached an acceleration sensor at the point displayed in the left of figure \ref{fig:TransPhoto} to measure an acceleration at the top of the detector. 
We already had a mock-up test on the transportation of the detector components such as PMTs and their support structures using the mini-truck with the same road course \cite{HinoJPS:2017}. The test was conducted in the more severe acceleration condition, and showed no damage on the detector components; therefore, the direct comparison of the acceleration measurements between the mock-up test and this stainless steel tank transportation can be done.

\begin{figure}[htbp]
    \centering 
    \includegraphics[width=.9\textwidth]{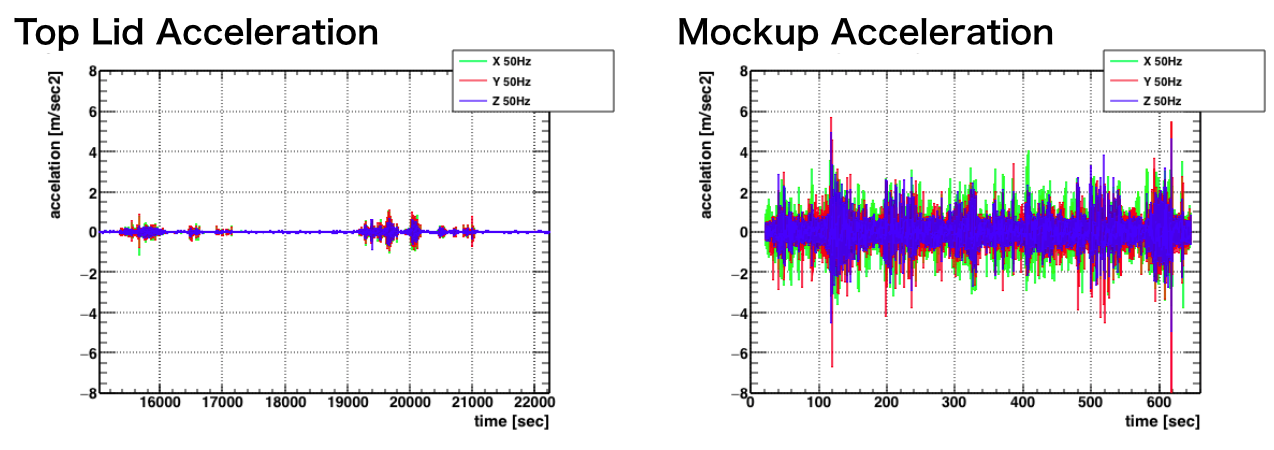}
    \caption{Plots of the acceleration sensor output. Left: the history of the acceleration on the top lid during transportation. Right: the data from the detector mock-up from \cite{HinoJPS:2017}. Note that the duration of the tank transportation were longer than that of the mockup test; however, it is obvious that the tank was exposed to smaller acceleration during the transportation compared to the mockup. \label{fig:TransPlot}}
\end{figure}

Figure \ref{fig:TransPlot} shows plots of acceleration during the transportation. The left plot is the result of the tank transportation, and the right one is that of the mock-up test, respectively.
The largest acceleration was around 1 \mssq during the tank transportation; in contrast, the mock-up had been exposed more than 1 \mssq for an entire duration of the round trip.
This result leads to a conclusion that all of the detector components get no hurt during the actual \JSNS detector transportation.

\section{Gas-tightness test \label{sec:gasleaktest}}
\begin{figure}[htbp]
    \centering 
    \includegraphics[width=.8\textwidth]{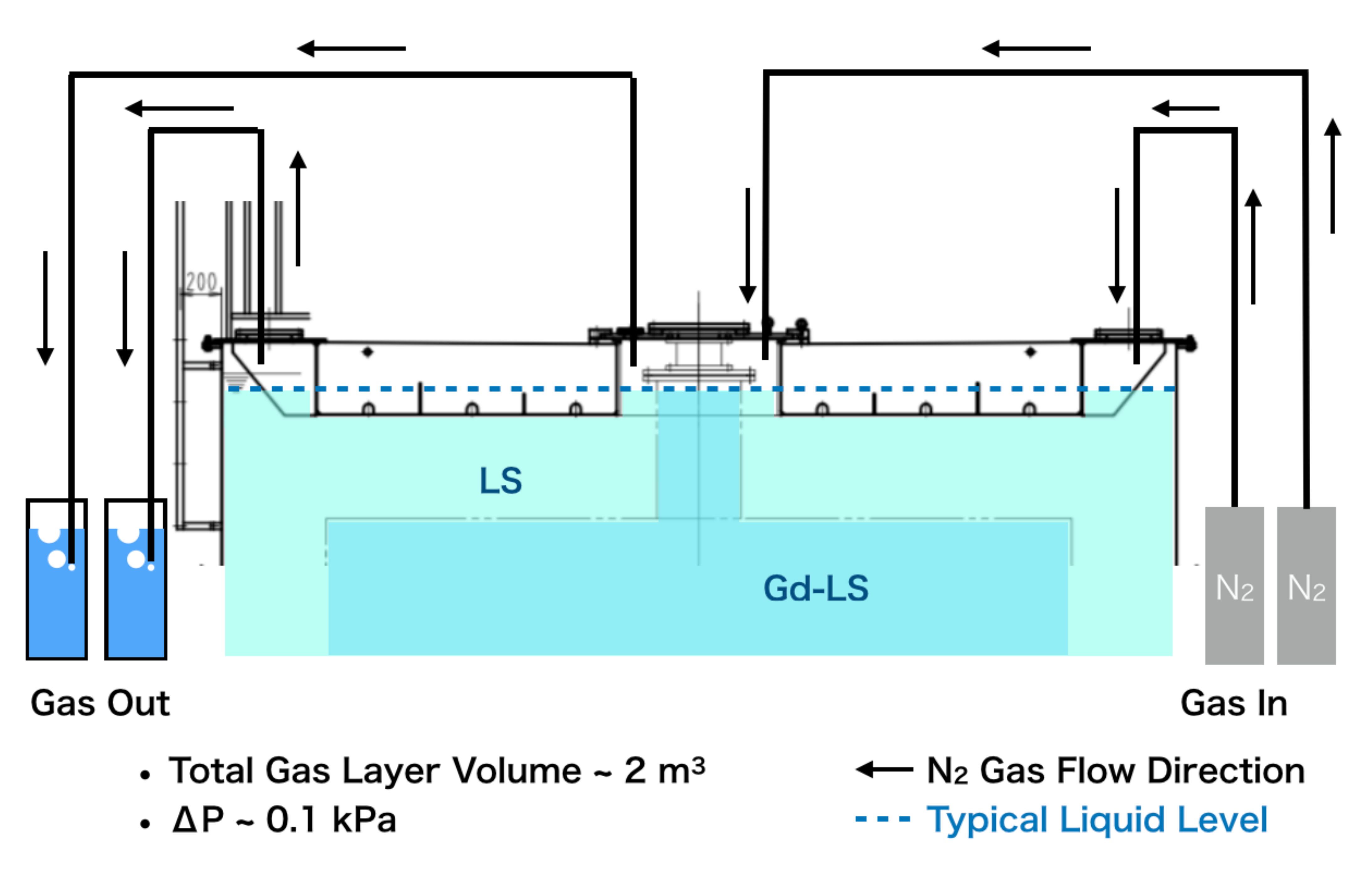}
    \caption{A concept of nitrogen gas flow system. Nitrogen gas is continuously supplied from gas cylinders to the gas phases of the \JSNS detector, and then goes to the bubblers. 
The end of the pipe of the nitrogen system is immersed 1 cm below the surface of the liquid to prevent air counter-flow. \label{fig:N2Flow}}
\end{figure}

As described above, to obtain a full performance of the IBD detection, it is crucial to keep the PSD capability in the entire volume of NT and GC for the fast neutron background rejection. Maintaining a nitrogen ambience for gas phases of the detector is essential to prevent oxygen contamination to Gd-LS and LS.
Figure \ref{fig:N2Flow} illustrates a concept of a bubbler system to keep a nitrogen ambience in the gas phases. Nitrogen gas is continuously supplied from gas cylinders to the gas phases of the ring buffer space and the chimney independently, and then goes 
to the bubbler. In the bubbler, the end of the pipe is immersed in the liquid. The depth is 1 cm below the liquid level, which is equivalent to about 0.1 kPa pressure difference between the atmosphere and gas in the tank. This nitrogen system prevents the air intrusion into the gas phase of the detector.
In order to keep the positive pressure,
it is important that the flanges on the detector have adequate sealing capability.
If the flow rate can be set to 100 mL/min, 
a nitrogen gas cylinder can supply nitrogen for 1.5 months.
Therefore, We set a tolerance level of total gas leakage as 100 mL/min, comparable amount to the nitrogen flow. 

\subsection{Concept of the test}
If the stainless steel tank contains higher pressure of gas than that of atmosphere, a relative pressure of the gas with respect to the atmospheric pressure ($\Delta P$) decreases as a function of time in case the tank has a leakage and no supplemental gas. The speed of the leak is proportional to $\Delta P$ at the moment. Therefore, the time evolution of $\Delta P$ follows an exponential function:

\begin{equation}
    \label{eq:leak_concept}
    \begin{split}
        \frac{\diff (\Delta P)}{\diff t} \propto - \Delta P \quad \Leftrightarrow \quad \Delta P(t) = \Delta P_0 \exp \left( -\frac{t}{\tau} \right) \,,
    \end{split}
\end{equation}

\noindent where $\Delta P_0$ is the relative pressure at $t=0$. This equation exhibits that the time constant $\tau$ characterize the leakage of the system. This corresponds to a time evolution of the number of gas molecules denoted as $n_{\mathrm{cor}}(t)$, that contributes to $\Delta P$.

\begin{equation}
    \label{eq:n_correspond}
    \begin{split}
        n_{\mathrm{cor}}(t) = n_{\mathrm{cor}}(0) \exp \left( -\frac{t}{\tau} \right) \,, \quad \mathrm{where} \quad n_{\mathrm{cor}}(0) = \frac{\Delta P_0 V}{RT} \,.
    \end{split}
\end{equation}

As 100 mL/min flow rate will be kept to maintain a nitrogen ambience during periods of a physics run of \JSNS, if we assume that a leak speed of the number of molecules equals that of the nitrogen flow, a time constant corresponding to the tolerance level of leakage is calculated as follows:

\begin{equation}
    \label{eq:n_leak}
    \begin{split}
        \left. \frac{\diff n_{\mathrm{cor}}(t)}{\diff t} \right|_{t=0} &= \frac{\Delta P_{\mathrm{g.p.}} V_{\mathrm{g.p.}}}{RT\tau} \sim \frac{1 \times 0.1 \left[ \mathrm{atm \cdot L/min}\right]}{RT} \,. \\
        \tau &\sim \frac{0.001 \times 2000}{1 \times 0.1} = 20 \, \left[ \mathrm{min} \right] \,,
    \end{split}
\end{equation}

\noindent where $\Delta P_{\mathrm{g.p.}}$, a relative pressure in the gas phases of the detector during physics runs, is kept about 0.1 kPa ($\sim 0.001$ atm) by the nitrogen flow system, and $V_{\mathrm{g.p.}}$ is total volume of the gas phases around 2 \mcub .

\subsection{Experimental Setup}
The gas-tightness test was done on June 2018. After closing the top lid with Herme-seal No.800 sealing, we supplied dry air and set up the relative pressure $\Delta P = 14.6$ kPa. To measure a relative pressure as a function of time, digital relative pressure sensor GC31 \cite{web:Nagano} with $\sim 0.5$ kPa resolution, was mounted on the one of small flanges on the top lid chimney. In addition to it, three different types of thermocoupples were used for monitoring temperatures of gas and the tank surface. Since a variation of atmospheric pressure affects the relative pressure measurement as well, an atmospheric pressure and temperature logging module TR-73U was placed outside of the tank \cite{web:TandD}. 
Analog outputs from each sensors, except for TR-73U, were acquired using data logger GL840 with 1/60 \Hz sampling \cite{web:Graphtec}. The measurement continued for about 3 days.

A time constant corresponding to the leak level of this setup is independent from $\Delta P$. However, the total volume of gas proportionally change the time constant. Because the gas volume of this setup is 30 times greater than that of the gas phases remaining after the detector is filled with Gd-LS/LS, e.g., during physic runs, a tolerance level time constant changes into

\begin{equation}
    \label{eq:tau_tolerance}
    \begin{split}
        \tau^{\mathrm{test}}_{\mathrm{tol}} \sim \tau^{\mathrm{ref}}_{\mathrm{tol}} \times 30 = 600 \, \left[ \mathrm{min} \right],
    \end{split}
\end{equation}

\noindent where $\tau^{\mathrm{test}}_{\mathrm{tol}}$ represents the time constant equivalent to the tolerance level of leakage in this setup, and $\tau^{\mathrm{ref}}_{\mathrm{tol}}$ is the time constant computed in eq. \eqref{eq:n_leak}.

\begin{figure}[htbp]
    \centering 
    \includegraphics[width=.8\textwidth]{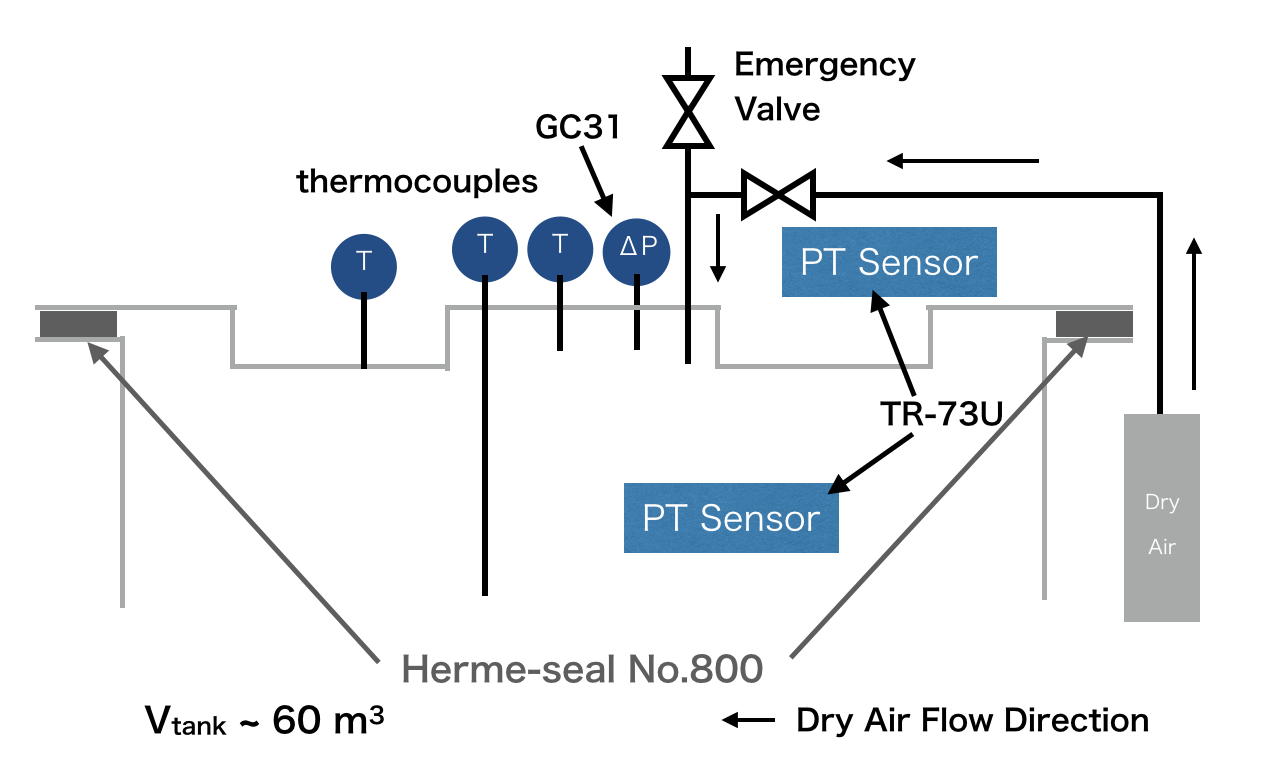}
    \caption{A schematic view of the gas-tightness test for the \JSNS detector. A relative pressure sensor and three thermocoupples were mounted on the top lid chimney flange. To monitor atmospheric pressure and temperature during the measurement, TR-73U sensor and logger module was used. Because the total volume is about 60 \mcub, a tolerance level of leakage corresponds to the time constant $\tau^{\mathrm{test}}_{\mathrm{tol}} \sim 600$ min in this setup. \label{fig:GasLeak}}
\end{figure}

The measured relative pressure data as a function of time is shown in figure \ref{fig:DataP}, where the horizontal axis represents the time interval from the beginning of the test. The markers in the plot indicate the relative pressure value at the moment. As a uncertainty from the resolution of the sensor, we assigned $\pm 0.5$ kPa systematic uncertainty to each point as vertical error bars.

\subsection{Prediction for Fit}
\begin{figure}[htbp]
    \centering 
    \includegraphics[width=.6\textwidth]{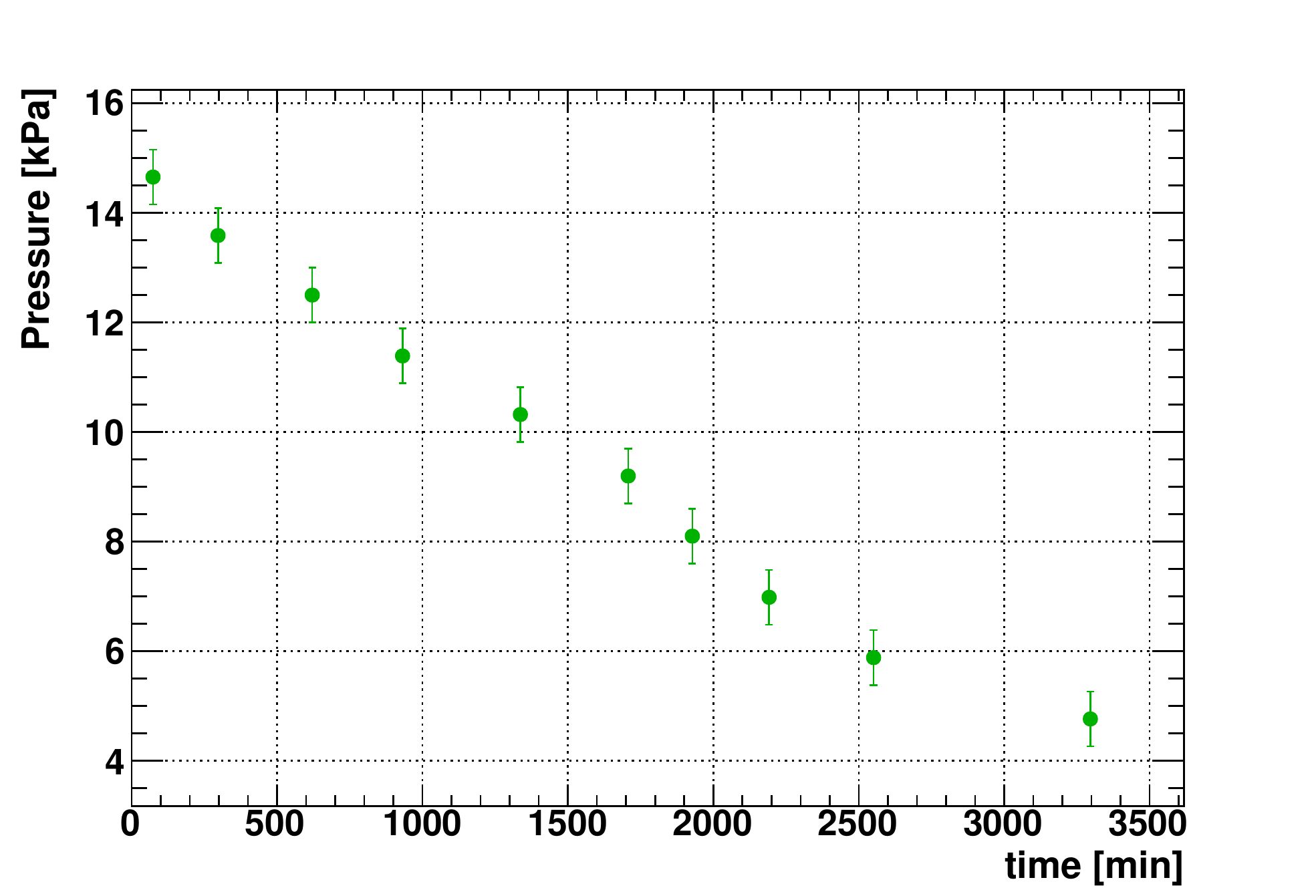}
    \caption{A plot showing the data of $\Delta P$ history during the measurement. The vertical axis shows the relative pressure with respect to atmospheric pressure in unit kPa. The origin of the horizontal axis corresponds to the time when we started the measurement. The vertical error bar comes from the resolution of the pressure sensor, and the horizontal one corresponds to the root mean square of each step. \label{fig:DataP}}
\end{figure}

\begin{figure}[htbp]
    \centering 
    \includegraphics[width=.9\textwidth]{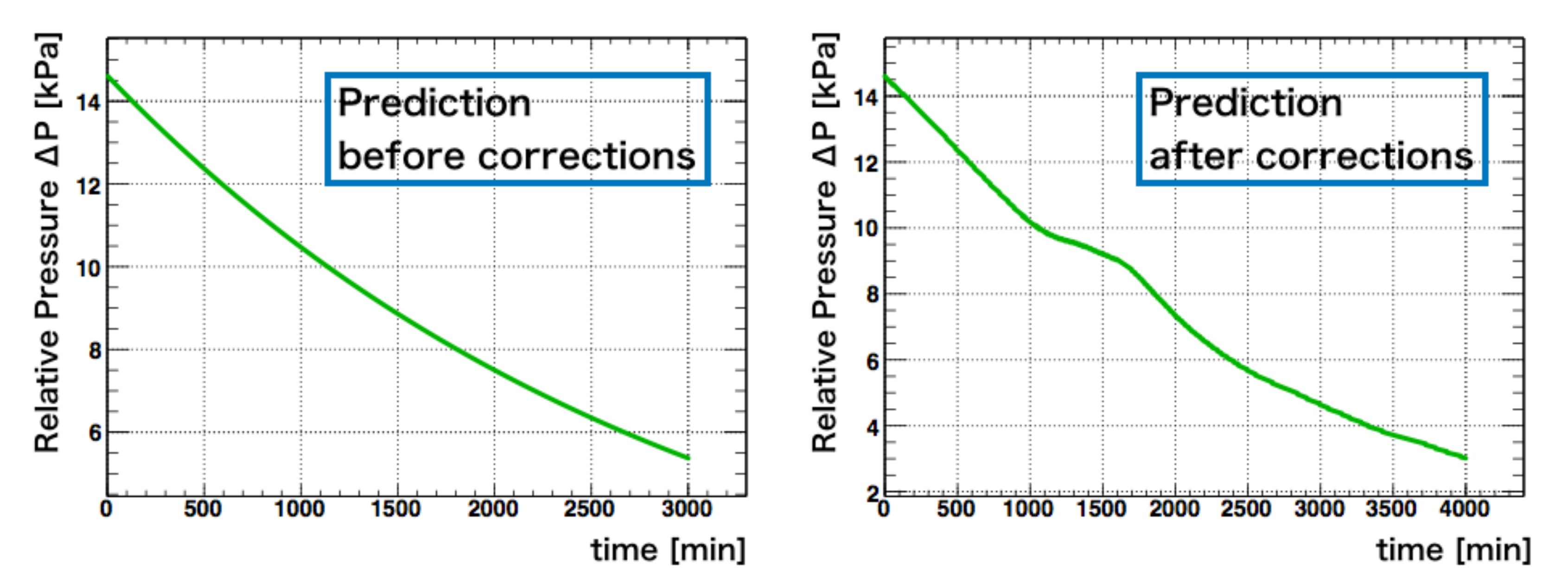}
    \caption{A comparison of predictions of $\Delta P$. Left plot shows the prediction without including the circumstance effects (computed based on eq. \ref{eq:euler_dp}). Right plot is the prediction including the environmental effects shown in figure \ref{fig:PatmTin}. $\tau = 3500 \mathrm{min}$ is assumed in both plots as an example. \label{fig:FitModel}}
\end{figure}

The change of environmental conditions, such as atmospheric pressure and gas temperature in the tank, causes a change of a relative pressure value on the sensor even though there is no leakage, and leads to a systematic error of the time constant measurement. 
Therefore, in order to extract the time constant from the data shown in figure \ref{fig:DataP}, we developed a prediction model simulating the relative pressure as a function of time, based on the supplemental data.
A discrete representation of eq. \ref{eq:leak_concept} can be shown as

\begin{equation}
    \label{eq:euler_dp}
    \begin{split}
        \Delta P^{n+1} = \left( 1-\frac{\Delta t}{\tau} \right) \Delta P^n \quad (n=0,1,\cdots) \,,
    \end{split}
\end{equation}

\noindent where $\Delta P^n$ is a relative pressure at each step, and $\Delta t$ shows a time interval corresponding to a step. Once an initial value $\Delta P_0$ is given, we can obtain time evolution prediction of $\Delta P$ successively (left of figure \ref{fig:FitModel}), which corresponds to Euler's method for numerical calculation of differential equations.

\begin{figure}[htbp]
    \centering 
    \includegraphics[width=.6\textwidth]{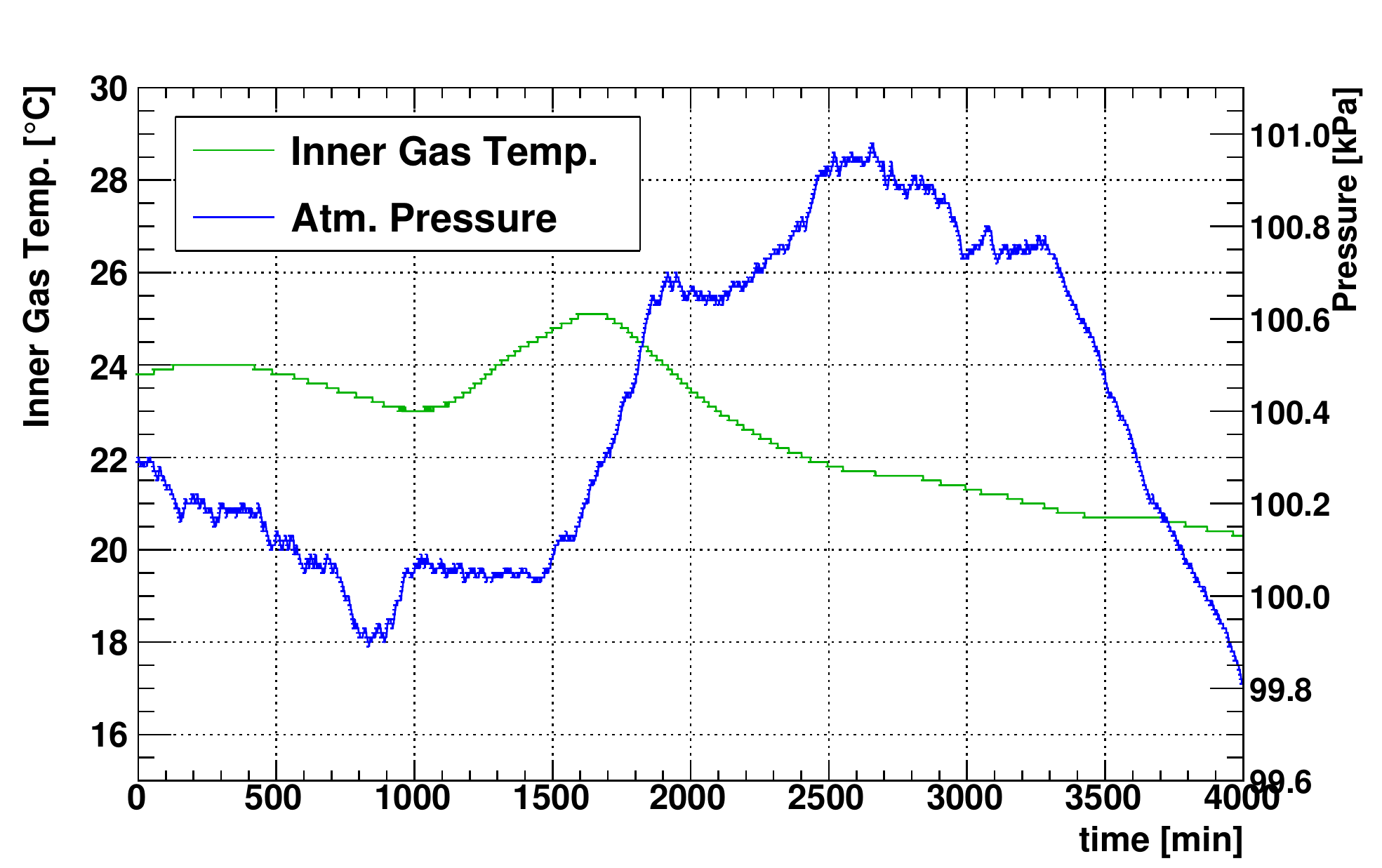}
    \caption{History of atmospheric pressure $P_{\mathrm{atm}}$ (blue) and the inner gas temperature $T_{\mathrm{in}}$ (green) in the tank. These data are used for corrections of $\Delta P$ prediction.  \label{fig:PatmTin}}
\end{figure}

To include the effect of the atmospheric pressure and the inner gas temperature change to the prediction, the logs of their data were used in an algorithm explained below.
\begin{itemize}
    \item First, a absolute pressure at the step $n$ is calculated using $P_{\mathrm{atm}}$ data shown as a blue line in figure \ref{fig:PatmTin}. A initial value of $\Delta P$ is set to the applied pressure $\Delta P_0$; therefore, $\Delta P^0 = \Delta P_0 = 14.6$ kPa.
        \begin{equation}
            P_{\mathrm{abs}}^n = \Delta P^n + P_{\mathrm{atm}}^{n} \,.
        \end{equation}
    \item An effect of a change of $T_{\mathrm{in}}$ is applied to $P_{\mathrm{atm}}$ on Boyle-Charles's law to obtain corrected relative pressure $\Delta P_{\mathrm{corr}}$. The inner gas temperature $T_{\mathrm{in}}$ was measured using the long thermocoupples shown in figure \ref{fig:GasLeak}, and exhibited as a light green line in figure \ref{fig:PatmTin}.
        \begin{equation}
            \begin{split}
                \Delta P_{\mathrm{corr}}^{n} &= P_{\mathrm{abs}}^{n} \times \frac{T_{\mathrm{in}}^{n}}{T_{\mathrm{in}}^{n-1}} - P_{\mathrm{atm}}^{n} \quad (n \ge 1) \,, \\
                \Delta P_{\mathrm{corr}}^{n} &= P_{\mathrm{abs}}^{n} - P_{\mathrm{atm}}^{n} \quad (n = 0) \,.
            \end{split}
        \end{equation}
    \item Finally, $\Delta P$ at the next step $n+1$ is obtained following eq. \eqref{eq:euler_dp} by substituting $\Delta P_{\mathrm{corr}}$ for $\Delta P$ to include the correction.
        \begin{equation}
            \Delta P^{n+1} = \left( 1 - \frac{\Delta t}{\tau} \right) \Delta P_{\mathrm{corr}}^{n} \,.
        \end{equation}
\end{itemize}

The right of figure \ref{fig:FitModel} displays the $\Delta P$ prediction including the environmental effects.
Note that $\Delta t$ represents 1 minute because they have data points in 1 minute interval. Both the circumstance data sets contain 4000 points of data. Thus, a maximum range of the prediction of $\Delta P$ can be obtained up to 4000 min, enough for fitting in entire range of $\Delta P$ data.

\subsection{Fit and Result}
\begin{figure}[htbp]
    \centering 
    \includegraphics[width=.6\textwidth]{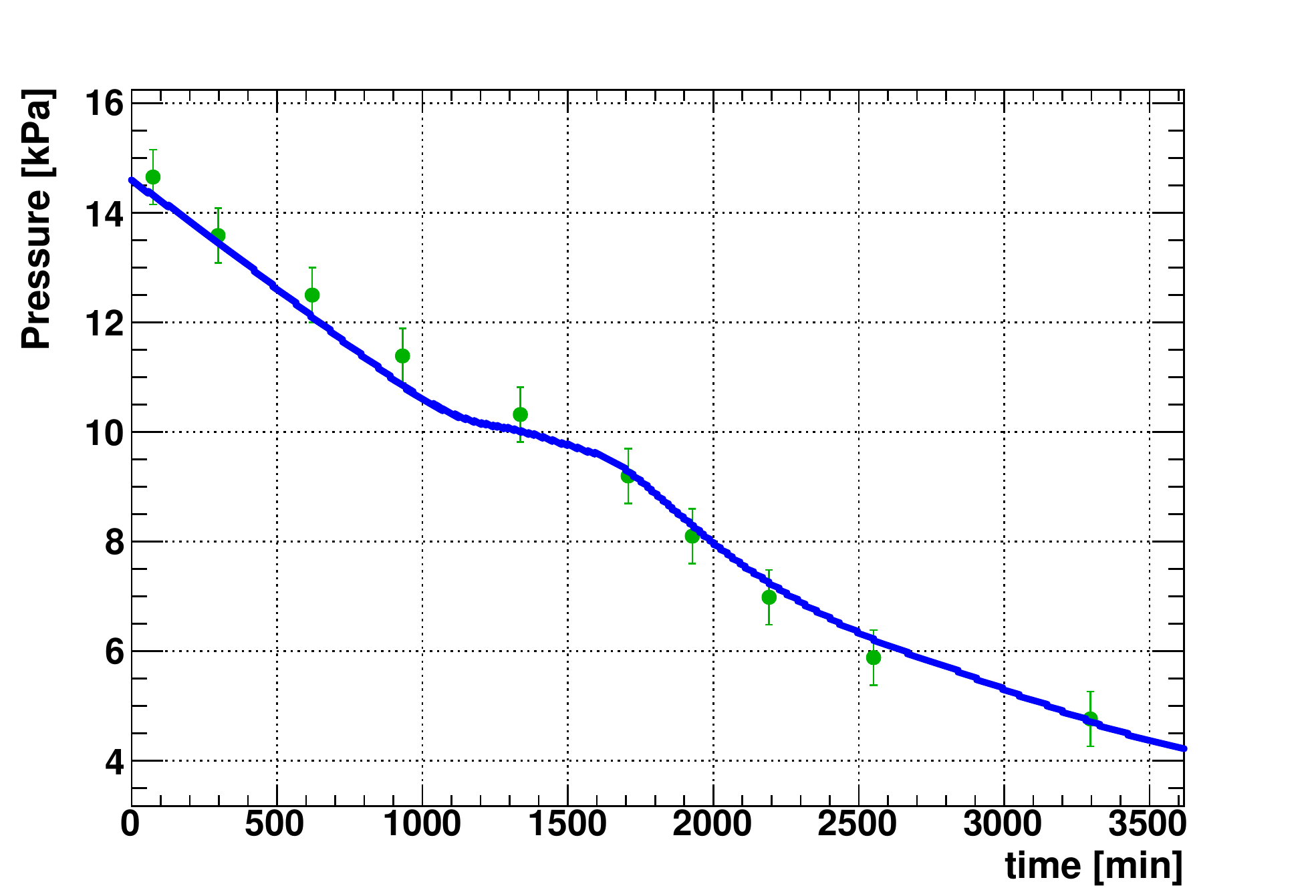}
    \caption{The result of the prediction fit showing the best fit curve (blue) at $\tau = 3410 \, \mathrm{min} \,$ with the processed data (green). \label{fig:BestFit}}
\end{figure}

Using the prediction including the correction with the circumstance data, we calculated $\chi^{2}$ with changing $\tau$ as a free fit parameter. The $\chi^2$ is calculated on a formulae

\begin{equation}
        \chi^2 = \sum_{i} \left( \frac{\Delta P^{\mathrm{data}}_i - \Delta P^{\mathrm{pred}}(t_i)}{\sigma^{\Delta P}_i} \right)^2 \, 
\end{equation}

\noindent where $\Delta P^{\mathrm{data}}_i \, \mathrm{and} \, t_i$ represent $i$-th data point in figure \ref{fig:DataP}, and $\sigma^{\Delta P}_i$ corresponds to their error bars respectively. As the prediction is computed in the discrete process explained above, a spline interpolation function denoted as $\Delta P^{\mathrm{pred}}(t_i)$ is used to get values between each point.
Figure \ref{fig:BestFit} shows a best fit curve of the prediction with $\chi^2/\mathrm{ndf} = 3.7/10$, where $\mathrm{ndf}$ is degree of freedom of the $\chi^2$. The error of $\tau$ is estimated as a value at $\Delta \chi^2 = 1$, where 

\begin{equation}
    \Delta \chi^2 = \chi^2 - \chi^2_{\mathrm{min}} \,.
\end{equation}

As a result of the fit, we obtained the time constant $\tau = 3410 \pm 144 \, \mathrm{min}$, whose central value and lower limit are more than 5 times larger than the tolerance level $\tau^{\mathrm{test}}_{\mathrm{tol}} = 600$ min.

\section{Summary}
The design, construction and tests for the stainless steel tank of the \JSNS detector have been reported.
The acceleration measurements and the mock-up test prove that the annual detector transportation between MLF and HENDEL can be done without problems.
We also developed a new sealing technique with the liquid gasket for a poor flatness flange, and applied to the connection flange of the tank. The capability of the sealing for gas was examined by observing a decrease in positive relative pressure as a function of time. As the result with a fit technique using the prediction of the pressure decline, we obtained over 5 times larger value of the time constant $\tau = 3410 \pm 144 \, \mathrm{min}$ than that of the tolerance level $\tau^{\mathrm{test}}_{\mathrm{tol}} = 600$ min.
These test results guarantees the stable performance of the \JSNS detector for a search for sterile neutrino in J-PARC MLF.


\acknowledgments
We warmly thank J-PARC and KEK for the various kinds of supports. We appreciate Morimatsu Industry Co., Ltd. for co-work as well.
This work is also supported by the JSPS grants-in-aid (Grant Number 16H06344, 16H03967), Japan.



\end{document}